\documentclass[12pt,
 reprint,
 amsmath,
 amssymb,
 aps,
]{revtex4}

\usepackage{braket}
\usepackage{graphicx}
\usepackage{dcolumn}
\usepackage{bm}
\usepackage{tikz}
\usepackage{amsmath}

\usepackage{chemformula}

\usepackage{setspace} 

\begin{document}

\preprint{APS/123-QED}

\title{Exact Entanglement correlation complements the chemical bond description}

\author{J\'ozef Spa\l{}ek}
\email{jozef.spalek@uj.edu.pl}
\author{Maciej Hendzel}
 \affiliation{Institute of Theoretical Physics, Jagiellonian University,\\ ul.~\L{}ojasiewicza 11, PL-30-348 Krak\'{o}w, Poland}

\date{\today}

\begin{abstract}
We analyze the properties of the exact solution obtained by us recently for the 
extended Hetiler-London model for chemical bonding which has an analytic form. The emphasis is put on defining two-particle entanglement correlation as the 
complementary characterization of the chemical bond and relating it to the partial 
atomicity and the so-called true covalency. The newly introduced characteristics remove the deficiency of the standard definition of covalency which now vanishes in the limit of the separated atoms. In effect, a gradual evolution of the system of two indistinguishable electrons in a bound state into their distinguishable correspondents can be traced systematically. The present analysis has a universal meaning and may also be applied to more complex systems.  
\end{abstract}

\maketitle

\clearpage
\newpage
A fully quantitative understanding of the nature of the covalent bond is fundamentally important for describing various chemical, physical, and biological systems. In the canonical case of the 
\ch{H2} molecule, the bond is depicted as a quantum mechanical bound state of two hydrogen atoms in the spin-singlet configuration of two electrons, minimally affected by the proton nuclear 
spins. Heitler and London established the initial quantitative description of the \ch{H2} molecule \cite{HeitlerLondon} within the essentially single-particle (Hartree-Fock) approximation of 
the two-electron state. This approach still serves as the foundation for a more comprehensive description, incorporating the mixing of virtually excited states of the constituent electrons 
\cite{PendasNature}.

Recently, we have rigorously solved the extended Heitler-London model, accounting for all-electron interactions in the ground state of the two-particle system with the 
inclusion of the \emph{1s}-orbital contraction contained in the resulting covalent and ionic parts of the total wave function \cite{Hendzel2}. A crucial feature of this exact solution is the 
derivation of the precise and analytic form of the two-particle wave function. This enabled a redefinition of the bond covalency and ionicity, along with the introduction of the degree of atomicity persistent upon bond formation, which in turn opens up a new path to a complementary bond characterization.

Here we demonstrate that the previously defined \emph{true covalency} \cite{Hendzel1, Broclawik, Hendzel3}, based on the ideas of the Mott-Hubbard localization, can be quantitatively 
related to the von Neumann entropy for the interacting and entangled two-electron states. Essentially, we name this entropy as \emph{the entanglement correlation} between the true covalent and 
ionic counterparts, thus complementing each other in the quantum mechanical manner. From this perspective, our findings complete the energy-based quantum mechanical description of the covalent bond, here 
exemplified for the homopolar molecule \ch{H2}. Moreover, our approach resolves the unphysical feature, observed in both the Heilter-London and the succeeding papers, for which 
the covalency increases with the increasing interatomic distance \cite{Pendas}. Such a refined analysis can be carried out in a physically clear way only by introducing the density matrix 
formulation with the admixture of atomicity to the pure bonding state. Finally, the quantum information point of view analysis provides us with an insight into the bond evolution with 
the interatomic distance $R$ gradually approaching the dissociation limit, $R \rightarrow \infty$, when the bonding electrons are getting localized on the parent atoms and thus become 
\emph{distinguishable in the quantum mechanical sense}.

\clearpage
\newpage
\textbf{THEORY.} We start with the expression for the two-particle wave function in the spin-singlet ground state, here rewritten in the second-quantized form \cite{Hendzel1}

\begin{align}
    \ket{\Psi_G} = \frac{2(t+V)}{\sqrt{2D(D-U+K)}} \left(
    \hat{a}^{\dagger}_{1\uparrow}\hat{a}^{\dagger}_{2\downarrow}-\hat{a}^{\dagger}_{1\downarrow}\hat{a}^{\dagger}_{2\uparrow}\right) \ket{0} \notag \\
     -\frac{1}{2}\sqrt{\frac{D-U+K}{2D}} \left( \hat{a}^{\dagger}_{1\uparrow}\hat{a}^{\dagger}_{2\downarrow}+\hat{a}^{\dagger}_{1\downarrow}\hat{a}^{\dagger}_{2\uparrow}\right) \ket{0} \label{wav1} \\
     \equiv 
     C\psi_{cov}(\textbf{r}_1,\textbf{r}_2) + I\psi_{ion}(\textbf{r}_1,\textbf{r}_2) \notag
\end{align}
where the first and second terms describe the covalent and ionic parts, respectively. The coefficients represent: $t$ and $V$ are the hopping and correlated hopping integrals, $U$ and $K$ are 
the magnitudes of the intraatomic and interatomic Coulomb repulsion, and $D \equiv \sqrt{(U-K)^2+16(t+V)^2}$. The creation operator part represents the intersite (first term) and intrasite spin songlet parts. They reflect microscopic parameters of the Hamiltonian \cite{Hendzel1, Broclawik} 
and are calculated microscopically within the EDABI procedure (see Methods). The representation Eq.~\eqref{wav1} is written in the basis of molecular wave functions. For the purpose of subsequent analysis we need a representation of Eq.~\eqref{wav1} in terms of single-particle Slater-atomic basis. Its explicit form is 
\begin{align}
    &\ket{\Psi_G} \equiv 
    \left(C\beta^2(1+\gamma^2) - 2\gamma I \beta^2\right)\ket{\phi^{at}_{cov}}  
    \notag \\ 
    &+\left( I\beta^2(1-\gamma^2) - 2\gamma C \beta^2 \right)\ket{\phi^{at}_{ion}} \notag \\
    &\equiv \tilde{C}\ket{\phi^{at}_{cov}} + \tilde{I}\ket{\phi^{at}_{ion}} \label{wav2}
\end{align}
where the sum $C^2+I^2$, as well as $\tilde{C}^2+ \tilde{I}^2$, should be both normalized to unity and their explicit form has been rigorously determined previously \cite{Hendzel1, 
Broclawik}. The physical meaning of the consecutive terms is the same as above. For the purpose of the present analysis, we write those components again in the explicit second-quantized form. Parenthetically, if we disregard the kets in Eq.~\eqref{wav1} the resulting entities represent the corresponding wave functions in the Schr\"odinger position representation.

One important side remark should be made at this point. Namely, when plotting the covalent and ionic coefficients with increasing interatomic distance $R$, then the covalency ($\gamma_{cov} \equiv C^2/(C^2+I^2)$) also increases and reaches its 
maximal value of unity in the $R \rightarrow \infty$ limit. This is an unphysical result and calls for a revision of the covalency definition in this form. To define \emph{the true covalency} ($\gamma_{cov} \equiv (\tilde{C}^2-\tilde{A}^2)/(\tilde{C}^2+\tilde{I}^2)$) we have introduced the atomic coefficient ($\tilde{A} \equiv C\beta^2$) and atomicity $\gamma_{at} \equiv \tilde{A}^2/(\tilde{C}^2+\tilde{I}^2)$ have subtracted the latter and then obtain the proper physical behavior of the quantities in the dissociation limit: $\gamma_{cov} \rightarrow 0$, $\gamma_{ion} \rightarrow 0$, and $\gamma_{at} \rightarrow 1$ \cite{Hendzel1, Broclawik}. 

By invoking the atomicity and hence defining the true covalency we introduced a mixed-state ingredient in the so-far pure-state quantum mechanical analysis. Therefore, to select a proper 
language for this new situation, we turn to the appropriate density matrix description. Intuitively, it amounts to replacing the covalent factor with its true covalency correspondent. Explicitly, 
as evident from Eq.~\eqref{wav1} we can define the two-particle density matrix in the form of $4x4$ matrix with the following trial states  $|\uparrow_1\downarrow_2\rangle$, $|\downarrow_1\uparrow_2\rangle$, $|\uparrow_1\downarrow_1\rangle$, and $|\downarrow_2\uparrow_2\rangle$ (where the subscripts $1$ and $2$ label the component atoms) from which all the spin-singlet states are composed of. In effect, the density matrix takes the starting form

\begin{align}
\rho = \frac{1}{C^2+I^2}\begin{pmatrix} C^2 & 0 & IC & 0 \\ 0 & C^2 & 0 & IC \\ IC & 0 & I^2 & 0 \\ 0 & IC & 0 & I^2 \\ \end{pmatrix} \label{eq:matrix}
\end{align}

Next, we subtract atomicity $\tilde{A}$ from the above $C$ factor in this expression. Then, the von Neumann entropy for this mixed state is determined in the following way 
\begin{align}
    S \equiv -Tr(\rho ln \rho) = -2(\gamma_{cov}+\gamma_{ion})ln(\gamma_{cov}+\gamma_{ion}) = 
    -2\gamma_{bond}ln(\gamma_{bond}). 
\end{align}
We are now ready to discuss the results and relate the bonding factors to the von Neumann entropy, which will allow us to interpret the latter as expressing the \emph{entanglement correlation}.

\begin{figure}[t]
    \centering
    \includegraphics[width=1.0\textwidth]{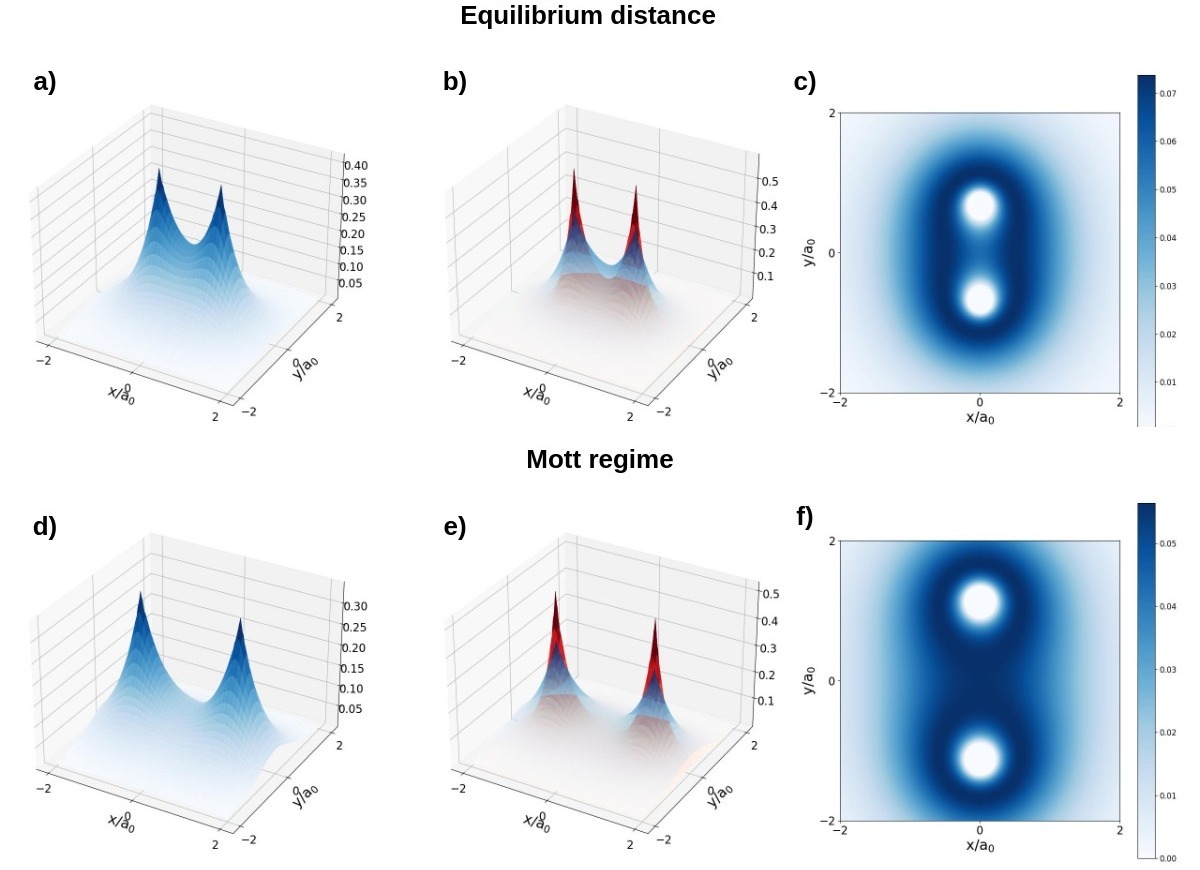}
    \caption{Left: Electron density profiles for two interatomic distances (equilibrium distance $R_{eq} = 1.43a_0$ and Mott-Hubbard boundary $R_{Mott} = 2.3a_0$) specified. Right: The purely covalent part of the wave function with the atomic and admixed ionic-covalent part subtracted. Note that the bonding is nonlocal and therefore requires a nonlocal (global characteristic): this is provided by either $\gamma_{cov}$ or $\gamma_{cov}+\gamma_{ion}$ or by the von Neumann entropy, as discussed in detail in the main text.}
    \label{fig:density}
\end{figure}

\textbf{RESULTS.} To illustrate the evolution of the bound state we have plotted in Fig.~\ref{fig:density} a-d electron density profiles, drawn 
on $(x,y)$ plane and at two interatomic distances: at equilibrium distance $R=R_{eq} = 1.43 a_0$ and the Mott-Hubbard threshold (when the kinetic-energy and interaction magnitudes 
are equal) $R=R_{Mott} = 2.3a_0$ \cite{Hendzel1, Broclawik}. The actual densities are shown in Fig.~\ref{fig:density}(a) and (d), whereas the red parts 
in (b) and (e) mark the atomic- and ionic-part contributions, while the blue parts mark the total electron 
density. Furthermore, Figs. (c) and (d) illustrate the corresponding densities on the plane 
$z=0$ with both the atomic part and the ionic-covalent admixture subtracted. From the 
two last profiles, one can see that the covalent bonding encompasses the space in 
between the two proton positions, which is not limited only to the region along the 
line connecting them, as one would presume intuitively, so although its overall 
magnitude is quite small. 

\begin{figure}[t]
    \centering
    \includegraphics[width=1\textwidth]{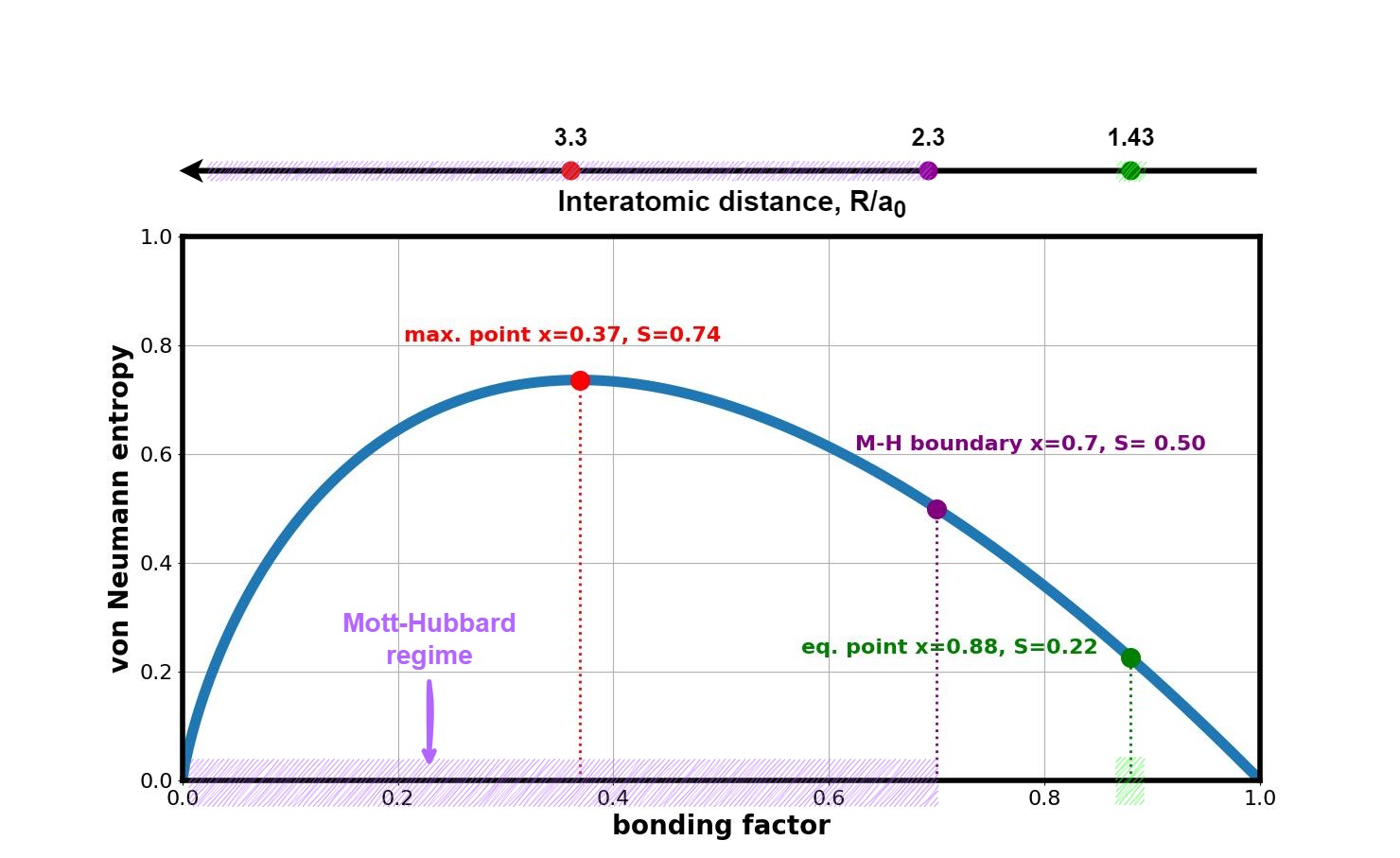}
    \caption{The von Neumann entropy as a function of bonding factor, $\gamma_{cov}+\gamma_{ion}$. The characteristic points are specified. The upper scale marks characteristic interatomic distances. The purple hatched area corresponds to the Mott-Hubbard regime, emphasizing the region where electronic correlations are predominant.}
    \label{fig:entaglement_covalency}
\end{figure}

In Fig.~\ref{fig:entaglement_covalency} we present the von Neumann entropy \emph{versus} the bonding factor, 
$\gamma_{cov} + \gamma_{ion}$ with characteristic points marked. On the upper scale the 
corresponding interatomic distances are labelled. The most important feature of this diagram is that the maximal entropy appears in neighborhood of the point where the kinetic energy and the interaction parts are comparable, i.e., close to the Mott-Hubbard (M-H) crossover point. The whole interaction-dominated M-H regime is also marked. Because of those features, we can call the entropy as the exponent of the \emph{entanglement correlations}. Note also that the entropy vanishes in the $R=0$ and $\infty$ limits, as we consider the dynamics only in the spin-singlet subspace of the total Fock space. The entanglement expresses the mutual correlation between the covalent and ionic parts induced by the competition part and the localization tendency induced by strong repulsive Coulomb interaction. The latter is responsible for the emergent atomicity.

The bonding features introduced by us as a complementary characterization to the standard 
energy considerations (see below) are summarized in Fig.~\ref{fig:entaglement_R}, all as a function of the interatomic distance. Explicitly, as said above, the point of maximal entanglement correlation is close 
to that, where the bonding $\gamma_{cov} + \gamma_{ion}$ and atomicity coincide. The latter 
may be regarded as a critical point for the crossover trend from the dominant single-particle 
bonding (characterized by the kinetic energy associated with the electron hopping) to the 
interaction-dominated (M-H) regime. The point is relatively far away from the equilibrium distance $R_{eq}$ but may play a crucial role in the molecule dissociation process under the 
presence of a catalyst. To summarize the meaning of Figs.~\ref{fig:entaglement_covalency} and~\ref{fig:entaglement_R}, they represent a fully complementary quantum characterization of 
the single chemical bond in the two-atomic molecules such as \ch{H2} and related systems \cite{Hendzel2}. Nonetheless, the homopolar \ch{H2} case represents the clearest situation as the atoms are the same and hence the quantum evolution with $R$, not distributed by extrinsic factors such as the component atoms inequivalence. 

\begin{figure}
    \centering
    \includegraphics[width=0.7\textwidth]{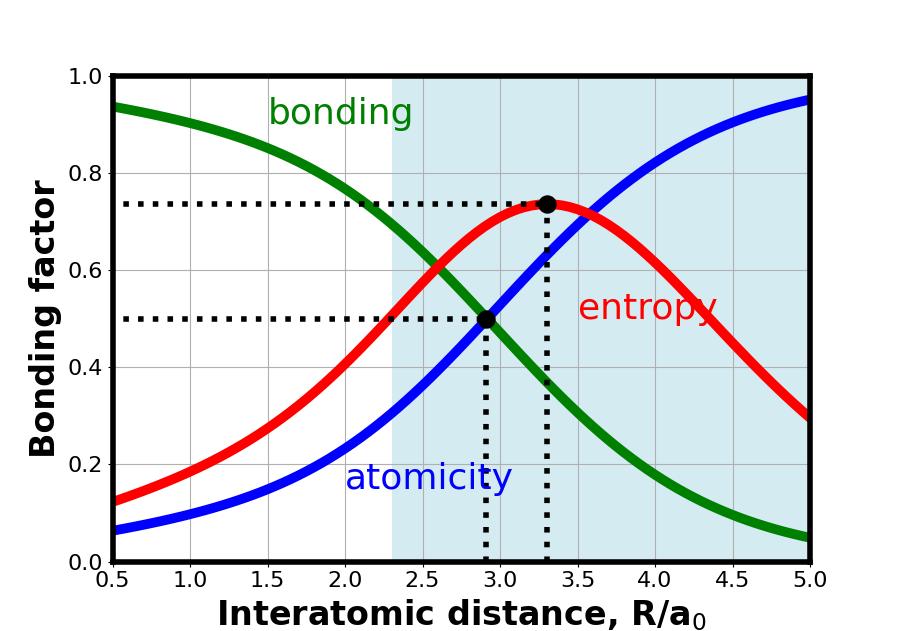}
    \caption{Relation between the bonding factor, atomicity and entanglement correlation (von Neumann entropy) as a function of interatomic distance. The shaded blue area corresponds to the Mott-Hubbard (crossover) regime \cite{Hendzel1, Broclawik}.}
    \label{fig:entaglement_R}
\end{figure}

\begin{figure}[t]
    \centering
    \includegraphics[width=0.7\textwidth]{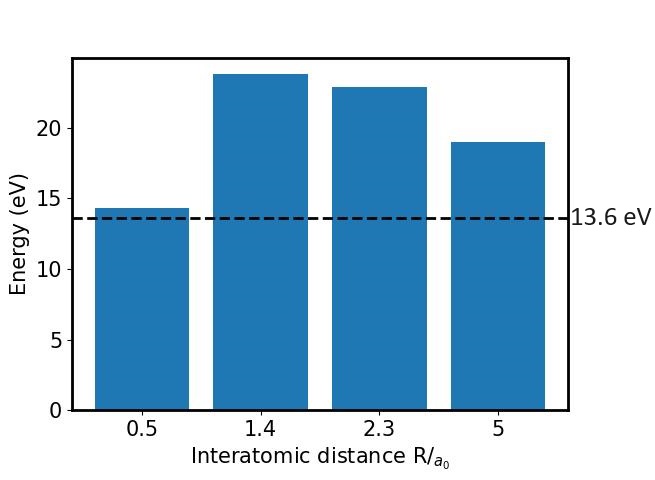}
    \caption{The adjusted atomic ($\epsilon_{at}$) dynamic contribution to the energy for \ch{H2} molecule with readjusted atomic orbital size for the specified interatomic distances $R$. The enhancement above reflects the orbital size contraction. The atomic energy is largely compensated by the intersite Coulomb interaction. }
    \label{fig:epsilon}
\end{figure}

\textbf{RELATED ENERGY CHARACTERISTICS.} 
\begin{figure*}[t]
    \centering
    \includegraphics[width=1.0\textwidth]{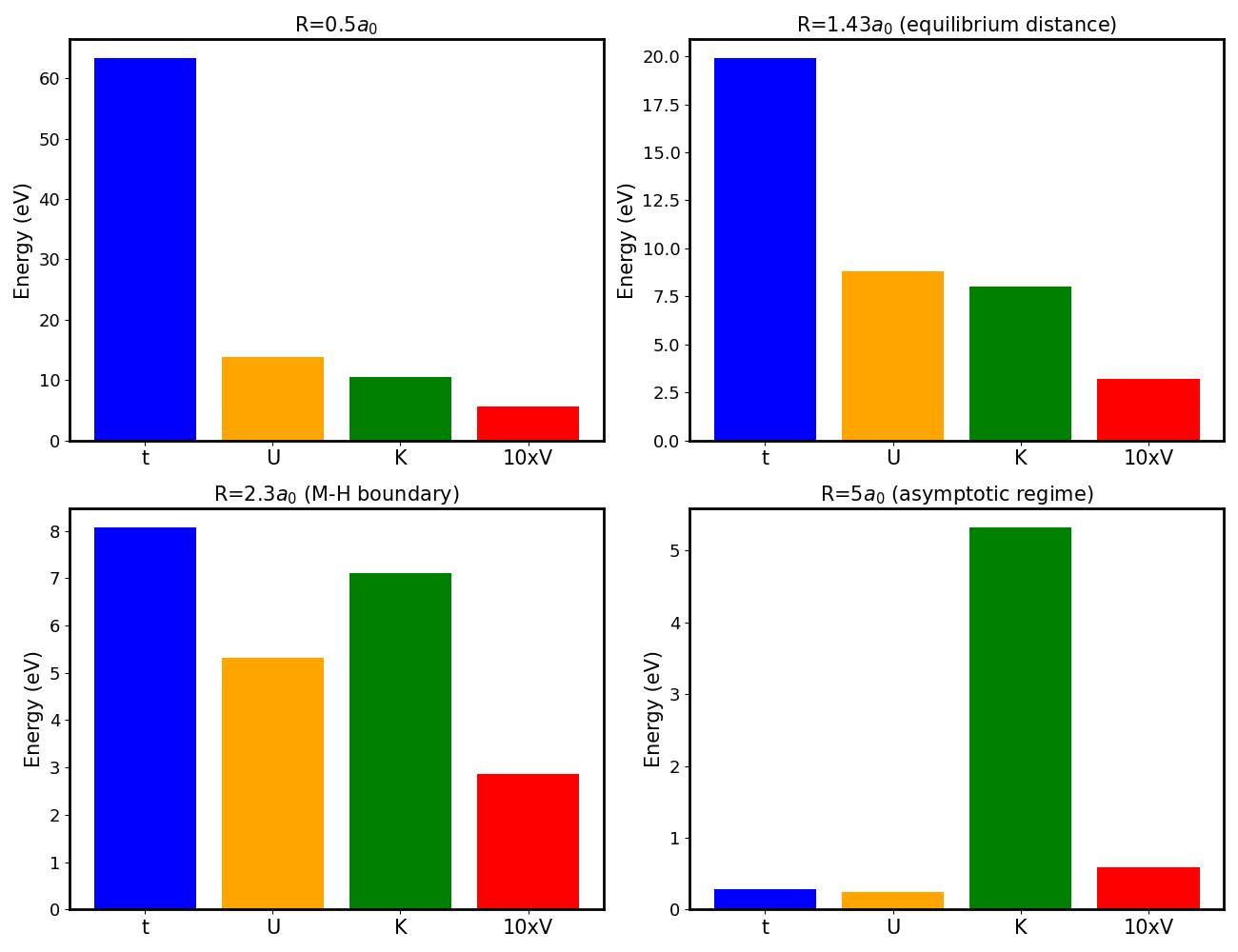}
    \caption{The bar charts specifying the relative energy contributions to the total energy of different components for four interatomic distance values ($1a_0$, equilibrium distance $1.43a_0$, Mott-Hubbard criterion $2.3a_0$, and $5a_0$). The Heisenberg exchange integral is not accounted for in the wave function coefficients and is disregarded here.}
    \label{fig:contributions}
\end{figure*}
Above we have connected directly the single-particle 
\emph{versus} interaction energy competition to the entanglement entropy related to it. As 
our method of approach allows for a detailed decomposition of the total bonding 
processes into various energy contributions, we now discuss on the evolution of 
the latter with the interatomic distance. In this manner, a comparison with a more 
standard approach can be made \cite{Zhao2019}. For that purpose, we plot in 
Fig.~\ref{fig:contributions} the bar charts of those terms composing the total ground state energy (per atom): hopping 
($2|t|\braket{\hat{a}^{\dagger}_{1\sigma}\hat{a}_{2\sigma}}$), intraatomic Coulomb interaction ($U\braket{\hat{n}_{1\uparrow}\hat{n}_{1\downarrow}}$) and interatomic ($\frac{1}
{2}K\braket{\hat{n}_1\hat{n}_2}$) Coulomb repulsion terms, and hopping correlation 
($2V\braket{\hat{n}_{1\uparrow}\hat{a}^{\dagger}_{1\downarrow}\hat{a}_{2\downarrow}}$) terms (the Heisenberg exchange term $\sim J$ has been ignored as it does not 
appear in the bonding factors). It follows that between $R=R_{eq}$ and $R=R_{Mott}$ the total hopping ($\sim 2|t+V|$) and the effective Coulomb interaction ($\sim U-K$) terms 
 are of comparable magnitude; this circumstance reflects the Hubbard criterion that ($2|t+V|/(U-K)) \simeq 1$ then. Parenthetically, as elaborated earlier \cite{Hendzel1, Broclawik} the criterion is the fundamental formal reference point of the whole analysis of atomicity. In effect, we cannot say whether the kinetic energy always represents 
the signature of the bonding. Instead, the factors displayed in Fig.~\ref{fig:epsilon} can be together unequivocally assigned as such. This conclusion can be illustrated further by 
calculating the dynamic correlations attached to the above terms. Namely, for example the orbital part $\braket{\hat{n}_1}+\braket{\hat{n}_2} = 2$, hopping probability 
$\braket{\hat{a}^{\dagger}_{1\sigma}\hat{a}_{2\sigma}} = 16|CI|$, intrasite Coulomb correlation $\braket{\hat{n}_{1\uparrow}\hat{n}_{1\downarrow}} = 4C^2$, intersite Coulomb correlation 
$\braket{\hat{n}_1\hat{n}_2} = 2I^2$, exchange spin correlation for the singlet state is obviously equal to unity, correlated-hopping correlation 
$\braket{\hat{n}_{1\uparrow}\hat{a}^{\dagger}_{1\downarrow}\hat{a}_{2\downarrow}} = 16|CI|$. In this manner, we relate directly the dynamic correlations to the bonding 
characteristics. This very important side conclusion could be also 
elaborated for heteropolar systems such as \ch{LiH}, \ch{HeH+}, etc., where the atomic energy part shifts the balance toward enhanced ionicity and atomicity. In the end, we should note 
that even the atomic part of the energy varies substantially with increasing $R$, as illustrated in Fig.~\ref{fig:epsilon}. We see that this energy (on an absolute scale) is enhanced in the molecular state with respect to the value of $R$ it would be in the limit of separated atoms (cf. dashed red line). The latest property provides an additional rationale for subtracting the atomic part in Figs.~\ref{fig:density}
(c) and (f) when explaining the meaning of the true covalency. 

\textbf{METHOD.} Our approach is based on the method EDABI (\textbf{E}xact \textbf{D}iagonalization \textbf{Ab I}nitio) 
proposed earlier for nanosystems with correlated electrons \cite{Hendzel1, Broclawik}. We start with the assumption of the trial
single-particle wave function basis which defines a truncated Hilbert-Fock space, in which a complete 
Hamiltonian in the second quantization representation is determined. The truncated basis is composed of both molecular (Mulliken-Wannier) states 
$w_{i\sigma}(\textbf{r}) \equiv w_{\sigma}(\textbf{r}-\textbf{R}_i)$, $i=1,2$ is formed in terms of which 
the microscopic parameters of the Hamiltonian $\mathcal{\hat{H}}$ in the Fock 
space are determined. The Wannier states are composed as a superposition of atomic Slater states 
with their size (determined analytically) the inter-
correlated state (eigenstates of $\mathcal{\hat{H}}$). The lowest eigenstate is one of the 
three \(^1 \Sigma_g^+\) ground states representing the ground state. 
The remaining five eigenstates (two singlets and 
three triplets) are the excited states. Finally, the two-particle 
wave function in an analytic form is determined by 
the transformation of the eigenvector in terms of the 
Fock space back to the Hilbert space \cite{JS2000}. The two-particle 
wave functions are determined analytically and for each 
of them, the crucial physical properties are discussed elsewhere [MH \& JS, in preparation].  

\textbf{OUTLOOK.} We reiterate that the molecular state, analyzed here within the exact solution of the extended Hetiler-London model, presents itself as a distinct case of the 
entanglement evolution when compared to that of noninteracting bosons \cite{HorodeckiReVMod}. This is because here the particles interact and for interatomic distance $R \rightarrow 
\infty$ gradually become disentangled but still bound to the parent atoms, i.e., \emph{distinguishable} in the quantum-mechanical sense \cite{Jordens2008}. In this respect the \emph{indistinguishability} $\rightarrow$ 
\emph{distinguishability} transformation is quite analogous to that of delocalization-localization (Mott-Hubbard) transition in condensed multiparticle systems \cite{Mott, Hubbard1964, PRLSpal}. Here the 
transformation is gradual but still conveys the same physics as the supercritical delocalization-localization behavior in the other nanosystems \cite{SpalJul}. In effect, it brings this aspect of Mott phenomenon to the 
molecular level. In effect, a degree of quantum statistical behavior may be implemented into the pure quantum-mechanical analysis of those systems. Obviously, the analysis of more 
involved and heterogenous systems (e.g. \ch{LiH}, \ch{HeH+}, \ch{H2-} etc.) may contribute to the possibility of modyfing the particle indistinguishability and analyze resulting specific 
emergent potentially atomic behavior.

One may also raise the question of how the behavior beyond the standard equilibrium configuration ($R=R_{eq}$) of molecules (e.g. \ch{H2}) can be made accessible for studying the most 
interesting situation with $R \gtrsim R_{Mott}$. One can think of screening the Coulomb interaction by invoking either its reduction by a medium or a proper catalyst. But then, one must be very careful and do not change the component particles' indistinguishability. Finally, the physics of, e.g., \ch{H2} molecule beyond $R=R_{eq}$ may be useful in enhancing its 
reactivity.

\section*{Acknowledgment}
This work was supported by Grants No.~UMO--2021/41/B/ST3/04070 and 2023/49/B/ST3/03545 
from Narodowe Centrum Nauki. The authors are very grateful to Ryszard, Michał, and 
Paweł Horodecki for their insights and critical remarks concerning the entanglement. We thank also our colleague Maciej Fidrysiak for numerous discussions. 

\clearpage
\newpage
\bibliography{hendzel}
\end{document}